\documentclass[aps,prb,twocolumn,showpacs,superscriptaddress]{revtex4}

\usepackage{graphicx} 

\begin{document}
\topmargin 0.25in

\title
{Pentagonal nanowires:
a first-principles study of atomic and electronic structure}

\author{Prasenjit Sen}
\affiliation{
Department of Physics,
University of Illinois at Chicago, Chicago, Illinois 60607-7059}
\author{O. G\"{u}lseren}
\affiliation{
NIST Center for Neutron Research,
National Institute of Standards and Technology,
Gaithersburg, Maryland 20899-8562}
\affiliation{
Department of Materials Science and Engineering,
University of Pennsylvania, Philadelphia, Pennsylvania 19104}
\author{T. Yildirim}
\affiliation{
NIST Center for Neutron Research,
National Institute of Standards and Technology,
Gaithersburg, Maryland 20899-8562}
\author{Inder P. Batra}
\affiliation{
Department of Physics,
University of Illinois at Chicago, Chicago, Illinois 60607-7059}
\author{S. Ciraci}
\affiliation{
Department of Physics, Bilkent University,
Ankara 06533, Turkey}

\date{\today}

\begin{abstract}

We performed an extensive first-principles study of nanowires in
various pentagonal structures by using pseudopotential plane
wave method within the density functional theory. Our results show
that nanowires of different types of elements, such as alkali,
simple, transition and noble metals and inert gas atoms, have a
stable structure made from staggered pentagons with a linear chain
perpendicular to the planes of the pentagons and passing through their
centers. This structure exhibits bond angles close to those in the
icosahedral structure. However, silicon is found to be energetically
more favorable in the eclipsed pentagonal structure.
These quasi one dimensional pentagonal nanowires have higher cohesive
energies than many other one dimensional structures and hence may be
realized experimentally.
The effect of magnetic state are examined by spin-polarized calculations.
The origin of the stability are discussed by examining optimized
structural parameters, charge density and electronic band structure,
and by using analysis based on the empirical Lennard-Jones type
interaction. Electronic band structure of pentagonal wires of
different elements are discussed and their effects on quantum ballistic 
conductance are mentioned. 
It is found that the pentagonal wire of silicon exhibits
metallic band structure.

\end{abstract}


\maketitle

\section{Introduction}

Very thin metal wires produced by the tip retracting from
nanoindentation in scanning tunneling microscopy (STM) or by 
mechanically controllable break junction (MCBJ)
have been subject of a number of experimental and theoretical
studies.\cite{gimzewski,agrait1,pascual1,krans,tgnini,frenken,review}
In particular, the step-wise behavior of the conductance measured in
the course of wire stretching at room temperature has attracted the
interests in various fundamental features of quantum theory, such as
the quantization of ballistic electron transport in very thin and one
dimensional conductors as well as Anderson's localization in very
long metal wires \cite{pascual2}. Recorded values of conductance just
before the breaking of the wire were in the range of the quantum of
conductance, $G_{o}=2e^{2}/h$. This implies that the smallest
cross sections of the wire are of atomic dimensions.
In fact, the conductance of suspended single atom gold wires,
which have been produced recently, is measured to be very close to
$G_o$ \cite{ohnishi1,yanson}. As pointed out 
earlier,~\cite{ciraci1,todorov,agrait2} force and
conductance variations measured  concomittantly during
stretching have indicated a close connection between the atomic
structure and the step-wise behavior of conductance. It is now understood that
a complex interplay between the quantization of electronic states 
with level spacing larger than room temperature, and the stable
structure having well defined number of atoms~\cite{stafford} and
also dynamic self-consistent potential in presence of a current
flow results in the observed step-wise behavior of conductance $G$ as
a function of stretching.

Apart from being a potential nanodevice with multiple operation modes
or ideal conducting connects between nanodevices, nanowires are
important because of their exotic and stable atomic structures
occurring in different sizes of cross sections. 
Therefore, a lot of effort is being devoted to the production of
nanowires that are conducting and stable. Metals crystallize in bulk
three dimensional structures because that is the most stable form. If
we wish to create one dimensional (1D) systems, there is clearly some
struggle against nature. Furthermore, 1D periodic metals
can suffer Peierls distortion and become non-metallic. For finite
nanowires, this tendency may be suppressed. However, for longer
nanowires we could end up with 1D systems that are either
unstable or insulating, both undesirable. Therefore, 
we should consider structural arrangements, 1D or
perhaps quasi-1D , which have cohesive energies as close
to the bulk as possible. Our search has led us to pentagonal
nanowires, a quasi-1D system, where a pair of pentagons
sandwich a single atom in a local icosahedral structure. The structure
looks like a pedestal lamp with a pentagonal base, is infinitely
repeated along the direction perpendicular to the planes of the
pentagons, is stable and does not suffer from Peierls distortion. We
performed extensive calculations and found that these pentagonal
quasi-1D nanowires have higher cohesive energies than many other 1D
structural arrangements.

The pentagonal structure
is incompatible with translational symmetry, and hence it is not
normally seen in 2D and 3D crystal structures. Strong evidence for five-fold
symmetric structures appeared in the first-principles molecular
dynamics simulations where the observations of a 13 atom stable
icosahedron of Na was reported.~\cite{barnett} The structure can be
viewed as a ``tiny pentagonal nanowire'' consisting of two pentagonal
bases with one Na atom present on either side of the pentagons. The
two pentagons share an apex Na atom and hence the Na$_{13}$ cluster.
Several composite structures with pentagonal motifs have also been
observed in simulated annealing study of ultra thin Al and Pb
nanowires.~\cite{oguz}
Subsequently, suspended monatomic chains,
strands and helical structures have been realized
experimentally.\cite{ohnishi1,yanson,ohnishi2} 
As the following discussion shows, part of the reason for the 
stability of the periodic pentagonal structures is that among several small
planar clusters, made of particles interacting through a two-body
Lennard-Jones potential, the pentagonal structure has the highest
binding energy per particle. 

Different regular
atomic structures occurring in different sizes are now a focus of
interest of experimental and theoretical studies seeking more
fundamental understanding of all these structures.
\cite{torres,maria,hakkinen,tolla,okamota,portal,prasen,bat,zhao}
Whether the pentagonal
structures predicted earlier by empirical
methods~\cite{mehrez,brand,oguz,hwang} are common to other elements
and can be understood from more fundamental principles have become
an important issue. In this paper we address this
question by using the first-principles plane
wave calculations within the density functional theory. We carry out
state-of-the-art total energy calculations for Na, Al, Cu, Au, Fe,
Ni, Pb, Si and Xe in two different pentagonal structures (four
structures for Au), and find
that the staggered pentagonal structure is a stable structure for
these elements except for Si. Furthermore, we compare the energetics
with other linear structures and perform an extensive analysis of
the electronic structure and charge density to reveal the origin
of stability and electronic properties of the pentagonal structure.
Finally, we mention the effect of the pentagonal structure on the
ballistic conductance.

\section{Description of the method and atomic structure}

First-principles plane wave calculations are performed within the
supercell geometry using a tetragonal unit cell. The axis of the wire
is taken along $z$-axis, and the lattice parameter of the wire
coincides with the lattice parameter $c$ of the tetragonal supercell.
The lattice parameters of the tetragonal cell in the x-y plane are
set as $a=b=15$ \AA~ so that the interaction between a wire and its
periodic images are negligible. Bloch states are expressed by the 
linear combinations
of plane waves with the cutoff energy $|{\bf k}+{\bf G}|^2$ always
larger than the optimum cutoff energy suggested for the ionic ultrasoft
pseudopotential~\cite{usps} of the element under study. The Brillouin
zone (BZ) integration is performed within Monkhorst-Pack
scheme\cite{monkh} using ($1 \times 1 \times 20$) ${\bf k}$ points.
Results are obtained by generalized gradient approximation~\cite{perdew}
(GGA). Preconditioned conjugate gradient (CG) method is used for wave
function optimization. To find the correct ground state we also performed
spin-polarized calculations for nanowires of Fe and Ni. Numerical
calculations are performed using both VASP \cite{vasp} and
CASTEP \cite{casteb} codes independently.

\begin{figure}[h]
\includegraphics[scale=0.35]{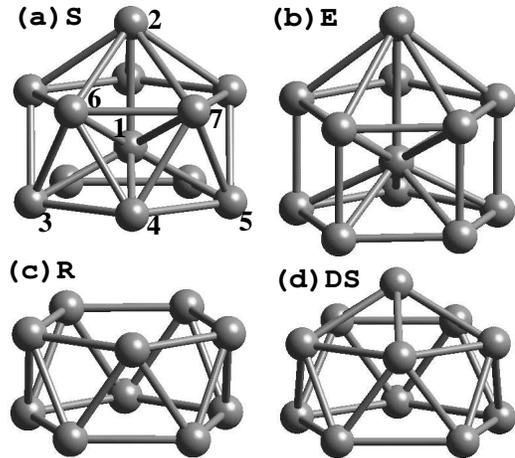}
\caption{Schematic description of various pentagonal structures
with the structural parameters: Lattice parameter along the wire, $c$,
and spacing between adjacent pentagons, $w$.
(a) $\mathcal{S}$ staggered pentagon structure with $c=2w$. Numerals
specify atoms. Atoms 1 and 2 form the chain passing through
the centers of the staggered pentagons. Relevant inter-atomic
distances
$d_{C-C}=w$, $d_{C-P}$, $d_{P-P}$ and $d_{P_{1}-P_{2}}$ are between
the atoms (1-2), (1-3), (3-4) and (4-6), respectively. Bond angles
$\alpha_{1}$, $\alpha_{2}$, $\alpha_{3}$, $\alpha_{4}$ and
$\alpha_{5}$ occur between (3-1-4), (3-1-5), (3-1-6), (3-1-7)
and (3-6-4). (b) In the eclipsed pentagon structure $\mathcal{E}$
all the pentagons are aligned. Hence $d_{C-C}=d_{P_{1}-P_{2}}=w$
and $c=w$ (c) Staggered pentagonal structure, $\mathcal{R}$, is
similar to $\mathcal{S}$ structure in (a), except that central atomic
chain is missing. $c=2w$ (d) The deformed staggered pentagon structure
$\mathcal{DS}$ with $c=2w$.}
\label{fig:struc}
\end{figure}

We considered the following pentagonal structures: (i) Atoms form parallel
pentagons which are perpendicular to the ($z$-) axis of the wire with
separation $w$, but successive pentagons are rotated by $\pi/5$. In
addition, a monatomic chain along the $z$-axis passes through the
center of pentagons, where each chain atom is located at a point
equidistant from the planes of pentagons. The 
lattice parameter $c$ in the direction of the chain
is twice the spacing between pentagons,
{\it i.e.} $c=2w$. This structure is specified
as the {\it staggered pentagon}, $\mathcal{S}$, and has 12 atoms in the
unit cell. (ii) Same as (i), but successive
pentagons are not rotated so that they have the same orientation
relative to the $z$-axis, and hence $c=w$. This structure is the {\it eclipsed 
pentagon}, $\mathcal{E}$, with 6 atoms per unit cell. 
(iii) Staggered pentagonal structure without 
the monatomic chain passing through the centers of the parallel pentagons, 
This is called the $\mathcal{R}$ structure and has $c=2w$. (iv) We
also found a 
modified version of the $\mathcal{S}$ structure in gold nanowires, 
which is specified as the {\it deformed staggered pentagon}, 
$\mathcal{DS}$. Here, adjacent pentagons are staggered, but one of 
the two atoms of the central chain in a unit cell is slightly displaced, while 
the other one is missing. Accordingly, $\mathcal{R}$ and 
$\mathcal{DS}$ structures have 10 and 11 atoms in a unit cell, 
respectively. The pentagonal structures and their relevant structural 
parameters are schematically described in Fig.~\ref{fig:struc}.

\section{Results and discussions}

\subsection{Optimized structures and energetics}

The energetics and atomic structure of Na, Al, Pb, Cu, Au, Ni, Fe,
Si and Xe wires in $\mathcal{S}$ and $\mathcal{E}$ have been
investigated. Our results for the structural parameters and binding
energies are listed in Table~\ref{table:pent}. The binding energies
and the relevant inter-atomic distances
corresponding to equilibrium bulk crystal structures are also given
for the sake of comparison. The binding energy per atom for a given
structure  is calculated as the difference of the energy $E_a$ of an
individual atom and the total energy of wire $E_T$ having $n$ atoms in
the supercell divided by $n$, {\it i.e.} $E_B=E_a - E_T/n$.
In spite of the fact that Na, Al, Cu, Pb, Au, Fe, Ni, Si, and Xe atoms
have different electronic configurations and form bulk crystals 
with dramatically different properties, they all form stable wires in
the pentagonal structure. It is very interesting that Al with $3s^2~3p$ 
valence states and Na with $3s$ valence state form similar 
pentagonal structures. Silicon, a group IV element which is normally 
crystallized in (tetrahedrally bonded) diamond structure is predicted 
to form pentagonal wires similar to what Xe, having a closed shell
structure,
does. The pentagonal structure is a stable structure
corresponding to a local minimum on the Born-Oppenheimer surface.

\begin{table*}
\caption{Comparison of calculated structural parameters and binding
energy, $E_{B}$, for different pentagonal structures of different
elements. The nearest neighbor distance, $d_{o}$, and binding
energy, $E_{0}$, are calculated for the optimized bulk crystals.
S, E, R, DS, and SM are staggered, eclipsed, staggered without central
chain, deformed staggered, magnetic staggered, respectively.
Bond lengths and energies are in \AA{} and eV respectively.}
\label{table:pent}
\begin{ruledtabular}
\begin{tabular}{c|c|cccc|ccccc|c||cc}
Atom & Structure & d$_{C-C}$ & d$_{C-P}$ & d$_{P-P}$ & d$_{P_1-P_2}$ &
$\alpha_1$ & $\alpha_2$ & $\alpha_3$ & $\alpha_4$ & $\alpha_5$ &
E$_B$ (eV) & d$_0$ & E$_0$ \\
\hline
Na & S & 3.02 & 3.75 & 4.1&3.67 &65 &114.5 &59 &114.5 &66 &1.054 &3.53 &1.28 \\
Na & E & 3.46&3.77 &3.95 &3.46 & 63&115 &55.3 &87 &47.3 &0.989 & & \\
Al & S & 2.39&2.71 &2.86 &2.82 &63.7 &117.2 &62.7 &116.2 &61 &3.201 &2.80
 &3.766 \\
Al & E & 2.54&2.76 &2.88 &2.54 &63.1 &115.3 &54.8 &88.5 &48.6 &3.189 & & \\
Al & R & & &2.68 &2.63 & & & & &61.3 &3.057 & & \\
Al & DS &4.54 &2.62 &2.77 &2.70 &63.9 &118 &62 &116 &58 &3.21 & & \\
Cu & S &2.21 &2.48 &2.61 &2.61 &63.4 &116.9 &63.5 &116.7 &60.2 &3.017 &2.58
 &3.76 \\
Cu & E &2.46 &2.48 &2.53 &2.46 &61.5 &111.3 &59.5 &91 &46 &2.878 & & \\
Pb & S &3.41 &3.29 &3.38 &3.45 &61 &111 &63 &114 &58 &3.18 &3.56 &3.51 \\
Pb & E &3.56 &3.32 &3.28 &3.56 &59.5 &106.5 &65.1 &93.5 &43 &3.13 & & \\
Au & S &2.50 &2.88 &3.05 &2.97 &63.8 &117.8 &62.4 &115.9 &61.7 &2.526 &2.95
 &3.211 \\
Au & E &2.76&2.87 &2.98 &2.76 &61.8 &113.7 &57.1 &89 &47.8 &2.494 & & \\
Au & R & & &2.89 &2.74 & & & & &63.4 &2.662 & & \\
Au & DS &4.51 &2.81 &3.15 &2.80 &68 &127.5 &58.2 &113.5 &69 &2.669 & & \\
   &    & &2.88 &2.98 &2.75 &62.5 &113.5 & & & & & & \\
Fe & SM &2.20 &2.50 &2.62 &2.61 &63.5 &117.2 &63 &117 &61 &7.298 &2.48 &8.37 \\
Fe & S &2.16 &2.31 &2.39 &2.51 &62.4 &114.2 &65.8 &117.5 &57 &6.563 & & \\
Ni & SM &2.19 &2.40 &2.51 &2.56 &63.1 &115.5 &64.4 &117.1 &58.7 &4.444 &2.49
 &5.48 \\
Ni & S &2.16 &2.40 &2.52 &2.53 &63.4 &116.5 &63.8 &117 &59.7 &4.337 & & \\
Ni & E &2.33 &2.38 &2.43 &2.33 &62.5 &112.3 &58.6 &89.8 &46.1 &4.20 & & \\
Si & S &2.62 &2.57 &2.60 &2.96 &60.7 &109.8 &70.1 &119.3 &52.2 &4.524 &2.35
 &5.40 \\
Si & E &2.73 &2.56 &2.55 &2.73 &59.6 &107.3 &64.2 &93.6 &43.1 &4.592 & & \\
Xe & S &3.74 &4.05 &4.23 &4.36 &62.9 &115.1 &65 &117.3 &58.1 &0.143 &
 4.51 & 0.06 \\
Xe & E &4.04 &4.05 &4.85 &4.04 &73 &142 &46.8 &81 &50 &0.104 & & \\
\end{tabular}
\end{ruledtabular}
\end{table*}

Calculated binding energies show that among the pentagonal
$\mathcal{S}$ and $\mathcal{E}$ structures the staggered one is 
energetically more favorable for Na, Al, Cu, Pb, Au, Fe, Ni and Xe.
However, the differences in binding energies,
$\Delta E_{B} = E_{B, \mathcal{S}} - E_{B, \mathcal{E}}$ are generally
small and are in the range of $\sim 10$ meV.
$\Delta E_{B} < 0 $ for Si, which favors the eclipsed
pentagonal structure. 

\begin{figure}[h]
\includegraphics[scale=0.35]{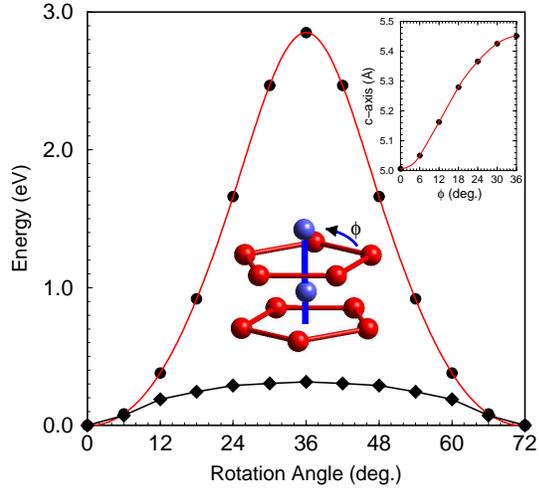}
\caption{ Variation of the total energy with the rotation angle
$\varphi$ from $\mathcal{S}$ structure ($\varphi=0$) to $\mathcal{E}$
structure ($\varphi=36^{o}$).
The maximum energy is the energy difference, $Q_{\mathcal{S}
\rightarrow \mathcal{E}}$, between the $\mathcal{S}$ and the
$\mathcal{E}$ structures.
Energies indicated by dots correspond to the robust rotation
of pentagons without relaxation of the structure. Energies indicated
by diamonds are calculated by fully optimizing the structure including
the lattice parameters at each step. The inset shows variation of
$c$.}
\label{fig:rot}
\end{figure}

We investigated the rotation between $\mathcal{S}$ and $\mathcal{E}$
structures of gold nanowires by breaking 
the rotation from $\mathcal{S}$ to $\mathcal{E}$ in seven
steps; at each step the relative angle ($\varphi$) between two
pentagons in a unit cell increased by an angle $\Delta \varphi = 6^o$.
For $0 \le \varphi < 36^{\circ}$ the lattice parameter $c$ is twice
the spacing between pentagons $w$, \textit{i.e.} $c=2w$.
The variation of the energy as a function of the rotation angle
$\varphi$ is illustrated in Fig.~\ref{fig:rot}.
The maximum of the energy curve corresponds to  
$Q_{\mathcal{S} \rightarrow \mathcal{E}}$. For robust rotation
of pentagons,  the wire is apparently under high compression due to the
core-core repulsion between two adjacent pentagons, and consequently
$Q_{\mathcal{S} \rightarrow \mathcal{E}}$ acquires a high
value $\sim$ 3 eV. However, upon optimizing the structure after
each rotation step, the structure is modified, in particular, the
spacing between adjacent pentagons increases with $\varphi$. As
a result, the energies calculated are lowered
dramatically, the curve for the variation of the energy as a
function of $\varphi$ is flattened and
$Q_{\mathcal{S} \rightarrow \mathcal{E}}$ is reduced to
$\sim$ 0.38 eV per cell (or 32 meV/atom).
We note that for 30$^{\circ} < \varphi <$ 42$^{\circ}$ around the
$\mathcal{E}$ structure the energy curve is practically flat.
This implies that $\mathcal{E}$ structure may acquire a helicity
along the axis of the wire, if each pentagon rotates by a small angle
$\varphi$. In fact, the helicity in nanowires was seen in
classical MD calculations~\cite{oguz} and has recently been observed
experimentally~\cite{ohnishi2}.
Nevertheless, it is clear that the $\mathcal{E}$ structure is
not only energetically less favorable, but is also unstable. 

From Table~\ref{table:pent} it is easily seen that this is the general
trend in all systems that show metallic bonding (Na, Al, Cu, Pb,
Au, Ni). Si also shows a similar behavior, but the relaxation of the
$z$ axis lattice constant in rotating the nanowire from the
$\mathcal{S}$ to the $\mathcal{E}$ structure is much less compared to
Na, Cu or Au. This is because, in a system with metallic bonding, the
electrons are largely delocalized and they screen the ion cores less
effectively. Hence, when the cores of the atoms in two successive
pentagons come closer on the structure being rotated from
$\mathcal{S}$ to $\mathcal{E}$, there is strong repulsion which tend
to increase $w$ and make the latter structure less favorable. For a
system like Si, showing directional bonds, there is much better
screening of the cores. In the $\mathcal{S}$ structure, each Si atom
in the pentagon forms four bonds--two bonds of 2.57 \AA~ with two
central chain atoms and two bonds of 2.60 \AA~ with two other atoms
in the pentagon. The third neighbor of a Si atom is 2.96 \AA~
apart, and there is no bond formation with it. On the other
hand, for a Si atom in the pentagon for the $\mathcal{E}$ structure,
there are 6 bonds--two bonds of 2.56 \AA~ with two central chain
atoms, two bonds of 2.55 \AA~ with two atoms in the pentagon and two
weaker bonds of 2.73 \AA~ with two Si atoms in the pentagons above and
below it. 

In course of structure optimization of gold wire, we found
two other structures--$\mathcal{R}$ and $\mathcal{DS}$ 
structures. (See Fig.~\ref{fig:struc}.) These structures have smaller 
numbers of atoms in their unit cells, and are found to be energetically 
more favorable than the $\mathcal{S}$ structure in case of gold nanowire.
Interestingly, $\mathcal{R}$ is found to be energetically less 
favorable than $\mathcal{S}$ for Al nanowires. The stability of 
the $\mathcal{S}$ structure is examined for transition metals Ni 
and Fe by performing both spin unpolarized and spin polarized 
calculations. We found that Ni and Fe nanowires in staggered
pentagon structures are stable for both nonmagnetic (spin unpolarized)
and magnetic (spin polarized) states. Although, for Ni, the energy
gain in the magnetic structure compared to the nonmagnetic one is small,
the spin polarized state of Fe staggered pentagonal wire increases the
binding
energy by 0.73~eV/atom. Similarly, the magnetic moment per atom is
also much larger in case of Fe ($\sim 3 \mu_B$) than in Ni
($\sim 0.77 \mu_B$).
 
The interatomic distance from the chain atom to the corners of a pentagon,
$d_{C-P}$, is slightly smaller than the nearest neighbor distance within 
the pentagon, $d_{P-P}$, but both distances are close to the nearest 
neighbor distance $d_{0}$ of the bulk crystal. Assuming that 
$d_{C-P} \approx d_{P-P}$, the coordination number for a chain atom is 
equal to 10, and that for an atom of the pentagon is 4. Hence, roughly
speaking, the 
average coordination number is 7. This is smaller than the bulk 
{\it fcc} and {\it bcc} coordination numbers 12 and 8, respectively. 
We note, however, the nearest neighbor distance between adjacent 
pentagons, $d_{P_{1}-P_{2}} \sim$ 14-20\% larger than $d_{C-P}$ and 
$d_{P-P}$. The bond angles, $\alpha_{1}$ - $\alpha_{5}$, are close 
to the bond angles of icosahedral structure, {\it i.e.} 63.4$^{\circ}$ 
and 116.6$^{\circ}$. Therefore, the local atomic configuration in the 
pentagonal wires mimics the icosahedral structure.\cite{oguz,ino}   

Calculated binding energies $E_B$ of pentagonal wires are lower than 
the calculated bulk binding energies $E_{0}$ in the last column of 
Table~\ref{table:pent}. This can be explained by higher coordination 
number in bulk crystals. On the other hand, the binding energies of 
pentagonal structure are higher than the binding energies of various 
monatomic chain structures. In Table~\ref{table:lin} the binding 
energies of $\mathcal{S}$ structure are compared with the binding 
energies of relevant monatomic chain structures calculated~\cite{prasen}
earlier for Al and Au. For example, three monatomic chain structures
of Al are linear ($L$), zigzag ($W$) and equilateral triangular ($T$)
chains have binding energies 1.87, 1.92 and 2.50 eV, respectively.
The coordination number of the $\mathcal{S}$ structure has an
intermediate value between those of 1D monatomic chain structures
and bulk crystal, so its binding energy, $E_{B}$=3.2 eV. Gold also
follows the same ordering. We note the general trend that the
binding energy increase with
increasing coordination number in different structures. This trend
is clearly observed in Table~\ref{table:lin} by going from the
monatomic linear chain to the bulk.
However, in the Xe pentagonal wires, the binding energy becomes larger than
the binding energy of the bulk fcc structure (excluding the contribution
of the Van der Waals interaction). On the other hand, all the nearest
neighbor distances ($d_{C-C},d_{C-P},d_{P-P}$, etc.) are much smaller
than the bulk nearest neighbor distance, $d_0$.

\begin{table}
\caption{ Comparison of nearest neighbor distance and binding energies 
of monatomic linear (L), zigzag (W), triangular (T) chain structures and 
bulk crystal (B) with the binding energy of the staggered
pentagon $\mathcal{S}$ structure calculated for Al and Au. The binding
energies of L, W and T structures are taken from Ref[\onlinecite{prasen}].}
\label{table:lin}
\begin{ruledtabular}
\begin{tabular}{l|cc|cc}
 & \multicolumn{2}{c}{Aluminum} &  \multicolumn{2}{c}{Gold} \\
 & $d$(\AA) & $E_b$(eV/atom) & $d$(\AA) & $E_b$(eV/atom) \\
\hline
L & 2.41 & 1.87 & 2.59 & 1.68 \\
W & 2.53 & 1.92 & 2.56 & 1.90 \\
T & 2.51 & 2.65 & 2.71 & 2.23 \\
S & 2.71 & 3.20 & 2.88 & 2.53 \\
B & 2.80 & 3.76 & 2.95 & 3.21
\end{tabular}
\end{ruledtabular}
\end{table}

\subsection{Energy band structure and total charge density}

\begin{figure}[h]
\includegraphics[scale=0.45]{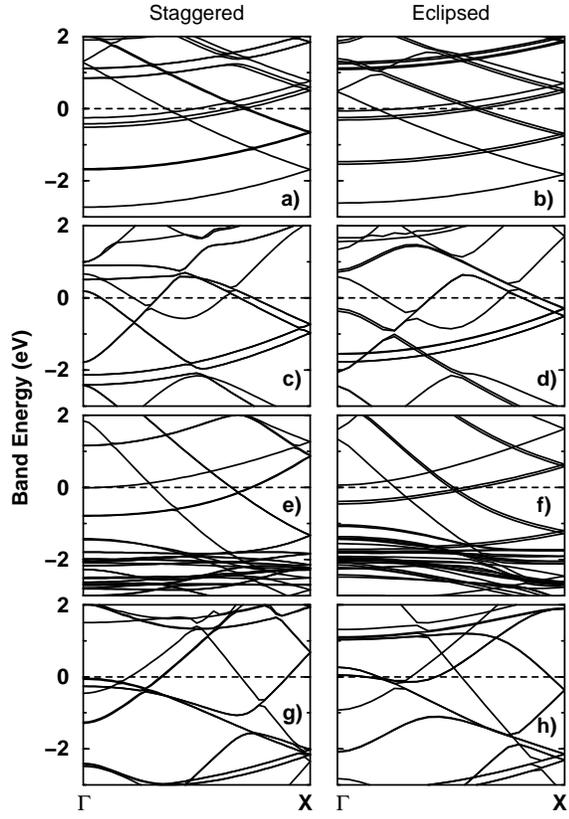}
\caption{ Energy band structures of $\mathcal{S}$ and $\mathcal{E}$
structures, respectively, (a) and (b) for Na, (c) and (d) for
Al, (e) and (f) for Au, and (g) and (h) for Si. Fermi level
shown by dashed lines marks the zero of energy.}
\label{fig:band}
\end{figure}

The electronic energy band structure of Na, Al, Au and Si are
presented for both $\mathcal{S}$ and $\mathcal{E}$ structures in
Fig.~\ref{fig:band}. Overall forms of the energy bands are
the same in both structures for Na, Al and Au, except for some 
shifts and splittings of degenerate bands. 
All these structures have the same number of atoms in their cells,
but Fermi level is crossed by different number of energy bands for
different elements. Number of bands which cross the Fermi level is
crucial for the quantum ballistic conductance and the stability of
nanowire. Under ideal conditions, the conductance
is determined by the number of bands crossing the Fermi level, being
$G_0$ per band. We found that 
the stable staggered pentagon structures of Na, Al and Au nanowires
have 6, 10 and 6 bands crossing the Fermi level, respectively.
On the other hand, the Fermi level of the unstable eclipsed pentagon
structure of Na, Al and Au nanowires is crossed by 6, 8 and
5 bands, respectively. While the staggered pentagonal structure
of Si nanowire have 6 bands crossing the Fermi level,
this number is raised to 10 for the eclipsed pentagon, which is the
stable structure. One notices that a degenerate band has dropped below
the Fermi energy in going from the $\mathcal{S}$ to the $\mathcal{E}$
structure for Si. This is possibly because of a weak bond
formation between Si atoms on different pentagonal planes, which is
absent in the $\mathcal{S}$ structure, and is responsible for the
$\mathcal{E}$ structure being more stable.
While stretching the nanowire, the number of atoms 
in the neck region, where the wire is thinned, and their structure 
exhibit sequential
and step-wise changes.\cite{mehrez,agrait2} It has been argued that
these changes are closely related with a band moving up from the
Fermi level and becoming unoccupied.\cite{stafford,ciraci2}
An important feature of Fig.~\ref{fig:band} is that the Si nanowire
is metallic in both structures. Several bands crossing the Fermi
level gives rise to high density of states at $E_F$. We found that
all pentagonal nanowires except Xe studied in this paper are metallic.
The Xe nanowire in $\mathcal{S}$ structure is a semiconductor with
a wide band gap.

\begin{figure}[h]
\includegraphics[scale=0.75]{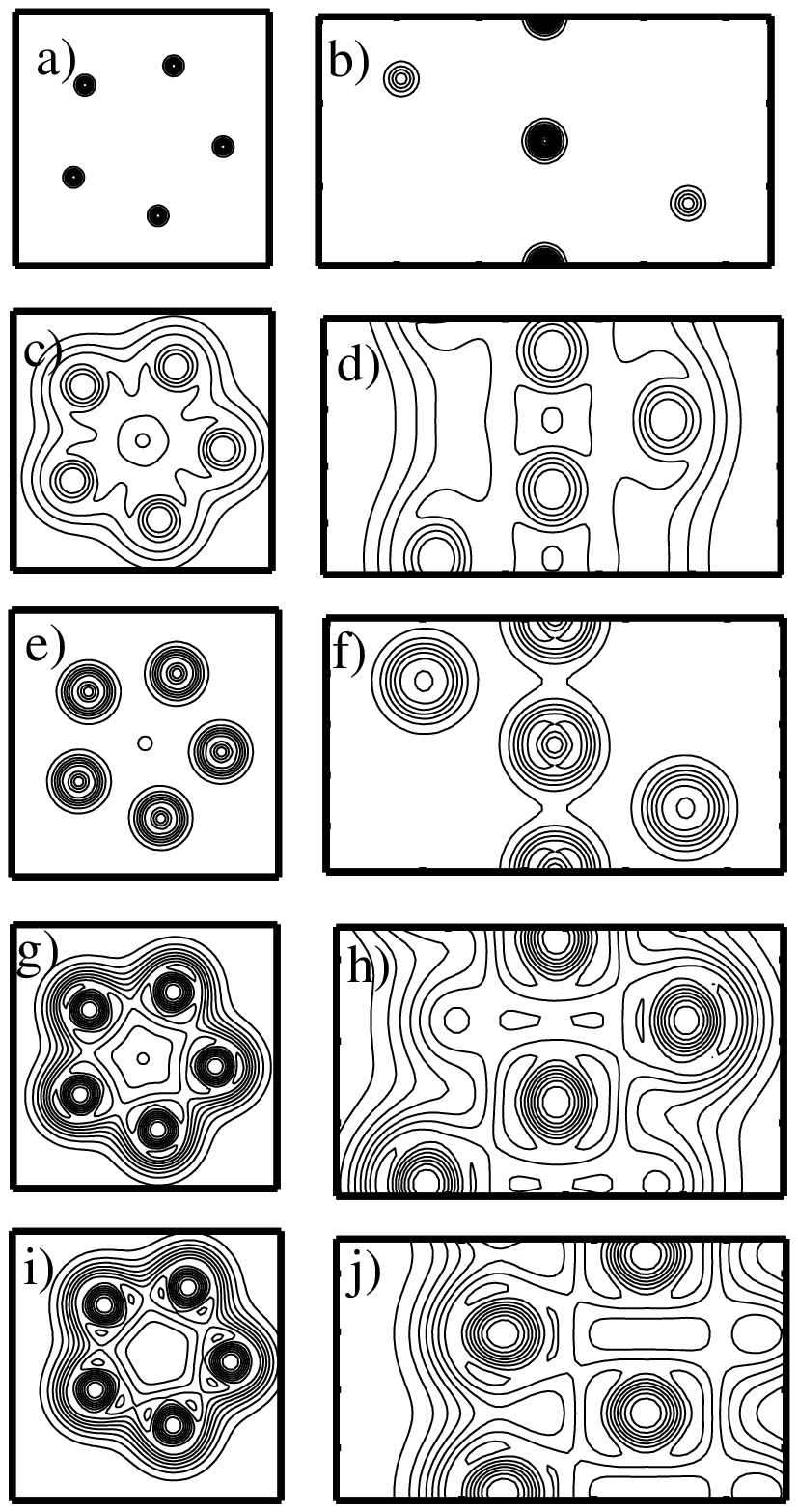}
\caption{ Left panels are charge density contour-plots in lateral
planes which coincide with the plane of pentagons. Right panels
are in vertical planes which pass through the central chain and
one atom of each pentagon. (a) and (b): Na in $\mathcal{S}$ structure.
(c) and (d): Al in $\mathcal{S}$ structure. (e) and (f): Au in
$\mathcal{S}$ structure. (g) and (h) Si in $\mathcal{S}$ structure.
(i) and (j): Si in $\mathcal{E}$ structure.}
\label{fig:char}
\end{figure}

The character of the bonding in pentagonal structures are revealed 
by the analysis of electronic charge density. In Fig.~\ref{fig:char}
we show the charge density contour plots of Na, Al, Au, and Si in 
different planes. The lateral plane passing through the plane of 
the pentagon show the character of the bonding between the  atoms
in a pentagon. The vertical plane includes the central chain as well
as one atom of the pentagon. Because of inversion symmetry of the
$\mathcal{S}$ structure, one atom of the adjacent pentagon is also 
included in the same vertical plane. The charge distribution
of Na is uniform in vertical and lateral planes. Because 
of low charge density significant structure cannot be resolved. 
The charge density contour plots in Fig.~\ref{fig:char}
display some differences in the charge distributions between different
atoms. The charge distribution of the central chain for Al is reminiscent
of that of monatomic chain structure\cite{prasen} and has a directional
character. The directional behavior is, however, less pronounced
in the bonds forming between two atoms in the same pentagon and also
between central chain atom and pentagon atom. The charge distribution 
of gold wire reflects the charge distribution of bulk metal. The 
directionality with high density along the line connecting two
nearest neighbor atoms is absent. Non uniform charge distribution with
a directionality between nearest neighbor atoms is clearly seen
in the contour plots of Si both in $\mathcal{S}$ and $\mathcal{E}$
structures. The directionality of charge distribution originates
from the valence of the element which makes the pentagonal wire.
Here, Al and Si with valence states consisting of $3s$, $3p_{x,y,z}$
orbitals form directional bonds. In contrast, Na and Au with
$s$ valence orbitals exhibit bulk like, uniform charge distribution
which is characteristic of metals. Despite these differences, all the
pentagonal nanowires in Fig.~\ref{fig:char} are metals with finite
density of states at the Fermi level.

\subsection{Discussions}

As corroborated by the present first principles calculations, the
pentagonal structure is one of the energetically favorable 
structures of wires having translational periodicity along its axis.
Of course, it is only a local minimum on the Born-Oppenheimer surface
and occurs for a given number of atoms in the cross section of the
wire (or in the 1D unit cell). Since the pentagonal structure occurs
for a number of different elements as demonstrated in this study,
the stability must stem from the pentagonal geometry. We examined
the relative stability of the pentagonal geometry by performing a simple 
analysis based on the two-body Lennard-Jones potential,
$V(r_{i,j})=4 \epsilon [(\sigma /r_{i,j})^{12}-(\sigma / r_{i,j})^{6}]$. 
We considered three structures encountered in the structure optimization
of very thin wires,\cite{mehrez,oguz} namely equilateral triangle, 
pentagon and hexagon. The total potential, 
$V_{T}=\sum_{i\neq j} V_{i,j}(r_{i,j})$ for these
structures are obtained by adding all the two-body potential energies,
$V_{i,j}$. By optimizing the total potential energy $V_T$ relative to
the nearest neighbor distance $d_{1}$, 
$\partial V_{T}/\partial d_{1}= 0$, the optimized interatomic
distances and the binding energies for these three structures are 
calculated in terms of
the parameters $\sigma$ and $\epsilon$. We found binding energies,
$E_{b}$ of $-\epsilon$, -1.1111 $\epsilon$, and -1.09 $\epsilon$,
for equilateral triangle, pentagon and hexagons, respectively.
The corresponding optimized nearest neighbor distances are
2$^{1/6} \sigma$, 1.89$\sigma$ and 1.91$\sigma$, respectively.
It is very interesting to note that among these structures, the 
pentagon has the highest binding energy. 

\section{Conclusions}

We carried out an extensive analysis of the energetics and structure of
thin wires made from different elements, such as alkali, simple, noble
and transition metals, and also Si, Xe. First-principles calculations
with fully optimized structures yield that the 1D structure
formed by parallel but staggered pentagons and an atomic chain passing
through the center of pentagons is generally stable and energetically 
favorable relative to other pentagonal structures. However, there are
exceptions. For example, while the eclipsed pentagonal structure
is favored by Si nanowires, in gold wires, different versions of pentagonal 
structures are found to be energetically more favorable. The binding energies
are intermediate between 1D chain structure and bulk crystal. 
All nanowires of different elements studied in this paper, except Xe, 
are metallic in the pentagonal structure. Strong cohesion and
metallicity of quasi-1D pentagonal nanowires suggest that they can be
useful in practical applications and deserve further experimental
studies.

\begin{acknowledgements} 
This work was partially supported by the NSF
under Grant No: INT01-15021 and T\"{U}B\'{I}TAK under Grant 
No: TBAG-U/13(101T010).
\end{acknowledgements}

\bigskip
{\bf Figure Captions}

\begin{thebibliography}{99}

\bibitem{gimzewski}
J. K. Gimzewski and R. M\"{o}ller,
Phys. Rev. B \textbf{41}, 2763 (1987).

\bibitem{agrait1}
N. Agra\"{i}t, J. G. Rodrigo, and S. Vieira,
Phys. Rev. B \textbf{47}, 12345 (1993).

\bibitem{krans}
J. M. Krans, C. J. M\"{u}ller, I. K. Yanson, T. C. M.
Gowarent, R. Hesper, and J. M. van Ruitenbeek,
Phys. Rev. B \textbf{50} (1994).

\bibitem{pascual1}
J. I. Pascual, J. Mendez, J. Herrero-Gomez, A. M. Baro, N. Garcia,
and V. T. Binh,
Phys. Rev. Lett. \textbf{71}, 1852 (1993).       

\bibitem{tgnini}
O. Tomagnini, F. Ercolessi, and E. Tosatti,
Surf. Sci. \textbf{287-288}, 1041 (1993).

\bibitem{frenken}
L. Kuipers, and J.W.M. Frenken,
Phys. Rev. Lett. \textbf{70}, 3907 (1993).

\bibitem{review}
A current review of the subject can be found in
S. Ciraci, A. Buldum, and I. P. Batra,
J. Phys.: Condens. Matter, \textbf{13}, R537 (2001).

\bibitem{pascual2}
J. I. Pascual, J. Mendez, J. Herrero-Gomez, A. M. Baro, N. Garcia,
U. Landman, W. D. Luedtke, E. N. Bogachek, and H. P. Cheng,
Science \textbf{267}, 1793 (1995).

\bibitem{ohnishi1}
H. Ohnishi, Y. Kondo, and K. Takayanagi,
Nature, \textbf{395}, 780 (1998).

\bibitem{yanson}
A. I. Yanson, G. R. Bolliger,  H. E. van der Brom,
N. Agra\"{i}t, and J. M. van Ruitenbeek,
Nature, \textbf{395}, 783 (1998).

\bibitem{ciraci1}
S. Ciraci and E. Tekman,
Phys. Rev. B \textbf{40}, R11969 (1989);
E. Tekman and S. Ciraci,
Phys. Rev. B \textbf{43}, 7145 1991).

\bibitem{todorov}
T. N. Todorov and A. P. Sutton,
Phys. Rev. Lett. \textbf{70}, 2138 (1993).

\bibitem{agrait2}
N. Agra\"{i}t, G. Rubio, and S. Vieira,
Phys. Rev. Lett. \textbf{74}, 3995 (1995);
G. Rubio, N. Agra\"{i}t, and S. Vieira,
Phys. Rev. Lett. \textbf{76}, 2302 (1996).

\bibitem{stafford}
C. A. Stafford, D. Baeriswyl, and J. B\"{u}rki,
Phys. Rev. Lett. \textbf{79}, 2863 (1997).

\bibitem{barnett}
R. N. Barnett and U. Landman,
Nature, \textbf{387}, 788 (1997).

\bibitem{oguz}
O. G\"{u}lseren, F. Ercolessi, and E. Tosatti, 
Phys. Rev. Lett. \textbf{80}, 3775 (1998).

\bibitem{ohnishi2}
Y. Kondo,and K. Takayanagi,
Science \textbf{289}, 600, (2000).


\bibitem{torres}
J.A. Torres, E. Tosatti, A. Dal Corso, F. Ercolessi,
J.J. Kohanoff, F.D. Di Tolla, and J.M. Soler,
Surf. Sci. \textbf{426}, L441 (1999).

\bibitem{maria}
L. De Maria and M. Springborg,
Chem. Phys. Lett. \textbf{323}, 293 (2000).

\bibitem{hakkinen}
H. H\"{a}kkinen, R. N. Barnett, A. G. Scherbakov, and U. Landman,
J. Phys. Chem. B \textbf{104}, 9063 (2000).

\bibitem{tolla}
F. D. Tolla, A. D. Corsa, J. A. Torres, and E. Tosatti,
Surf. Sci. \textbf{454-456}, 947 (2000).

\bibitem{okamota}
M. Okamoto and K. Takayanagi,
Phys. Rev. B \textbf{60}, 7808 (1999).

\bibitem{portal}
D. Sanchez-Portal, E. Artacho, J. Junquera, P. Ordejon,
A. Garcia, and J.M. Soler,
Phys. Rev. Lett. \textbf{83}, 3884 (1999).

\bibitem{prasen}
P. Sen, S. Ciraci, A. Buldum, and I. P. Batra,
Phys. Rev. B \textbf{64}, 195420 (2001).

\bibitem{bat}
I. P. Batra, P. Sen and S. Ciraci,
J. Vac. Sci. and Technol. (To appear in the May/June 2002 issue).

\bibitem{zhao}
B.L. Wang, S.Y. Yin, G.H. Wang, A. Buldum, and J.J. Zhao,
Phys. Rev. Lett. \textbf{86}, 2046 (2001);
B.L. Wang, S.Y. Yin, G.H. Wang, and J.J. Zhao,
J. Phys.: Condens. Matter, \textbf{13}, 403 (2001).

\bibitem{mehrez}
H. Mehrez and S. Ciraci,
Phys. Rev. B \textbf{56}, 12632 (1997);
H. Mehrez, S. Ciraci, A. Buldum and I. P. Batra,
Phys. Rev. B \textbf{55}, R1981 (1998). 

\bibitem{brand}
M. Brandbyge, K. W. Jacobsen, J. K. Norskov,
Phys. Rev. B \textbf{55}, 2637 (1997);
{\it ibid} \textbf{56}, 14956 (1997).


\bibitem{hwang}
J. W. Kang and H. J. Hwang,
J. Phys.: Condens. Matter, \textbf{14}, 2629 (2002).

\bibitem{usps}
D. Vanderbilt,
Phys. Rev. B \textbf{41}, 7892 (1990).
 
\bibitem{monkh}
H.J. Monkhorst, and J.D. Pack,
Phys. Rev. B \textbf{13}, 5188 (1976).

\bibitem{perdew}
J. P. Perdew, and Y. Wang,
Phys. Rev. B \textbf{46}, 6671 (1992).

\bibitem{vasp}
G. Kresse and J. Hafner,
Phys. Rev. B \textbf{47}, R558 (1993);
G. Kresse and J. Furtm\"{u}ller,
{\it ibid.} \textbf{54}, 11169 (1996).

\bibitem{casteb}
M.C. Payne, M. P. Teter, D. C. Allen, T. A. Arias
and J. D. Joannopoulos,
Rev. Mod. Phys. \textbf{64}, 1045 (1992).

\bibitem{ino}
S. Ino,
J. Phys. Soc. Jpn. \textbf{27}, 941 (1969).

\bibitem{ciraci2}
S. Ciraci and I.P. Batra,
Phys. Rev. B \textbf{33}, 4294 (1986);
I.P. Batra, S. Ciraci, G.P. Srivastava, J.S. Nelson,
and C.Y. Fong,
Phys. Rev. B \textbf{34}, 8264 (1986).

\end{thebibliography}
\end{document}